\documentclass{amsproc}

\usepackage{}
\usepackage{graphicx}
\usepackage{amsmath, amssymb, graphics,bbm}
\usepackage[justification=centering]{caption}
\usepackage{subfig}

\newtheorem{theorem}{Theorem}[section]

\theoremstyle{definition}

\theoremstyle{remark}

\numberwithin{equation}{section}

\begin{document}

% \title[short text for running head]{full title}
\title{Polynomial Rings and Topological Strings}

%    Only \author and \address are required; other information is
%    optional.  Remove any unused author tags.

%    author one information
% \author[short version for running head]{name for top of paper}
\author{Murad Alim}
\address{Department of Mathematics, Harvard University, 1 Oxford Street, Cambridge MA, 02138, United States}
\curraddr{}
\email{alim@math.harvard.edu}
\thanks{I would like to thank Dominique L\"ange, Peter Mayr, Emanuel Scheidegger, Shing-Tung Yau and Jie Zhou with whom I have collaborated on projects related to the current exposition.}

%    author two information
%\author{}
%\address{}
%\curraddr{}
%\email{}
%\thanks{}

%    The 2010 edition of the Mathematics Subject Classification is
%    now available.  If you are citing a classification from the
%    new scheme, use the following input coding instead.
%\subjclass[2010]{Primary }

\date{}

\begin{abstract}
An overview is given of the construction of a differential polynomial ring of functions on the moduli space of Calabi-Yau threefolds. These rings coincide with the rings of quasi modular forms for geometries with duality groups for which these are known. They provide a generalization thereof otherwise. Higher genus topological string amplitudes can be expressed in terms of the generators of this ring giving them a global description in the moduli space.  An action of a duality exchanging large volume and conifold loci in moduli space is discussed. The connection to quasi modular forms is illustrated by the local $P^2$ geometry and its mirror, the generalization is extended to several compact geometries with one-dimensional moduli spaces.
\end{abstract}

\maketitle

%%%%%%%%%%%%%%%%%%%%%%%%%%%%%
\section{Motivation and summary}

\subsection{Motivation}
The study of physical theories in families can be mapped to the study of deformation families of geometries. This has lead to a very fruitful interaction between mathematics and physics and to deep insights into both fields. The study of deformation families of superconformal algebras lead for example to mirror symmetry. Mirror symmetry identifies mirror families of Calabi-Yau (CY) threefolds where the moduli space $\mathcal{M}$ of the family on one side the that of comlpexified K\"ahler forms of a CY $X$, on the other side it is the moduli space of complex structures of the mirror CY threefold $\check{X}$.\footnote{Further background on mirror symmetry and references can be found in Ref.~\cite{Alim:2012gq}}

%%%%%%%%%%%%%%%%%%%%%%%%%%%%%%%%

\subsection{Summary}
Using the special geometry of $\mathcal{M}$, a special set of functions can be defined which form a differential ring, i.~e.~they close under derivatives. This ring was first put forward for the quintic and related Calabi-Yau geometries with one-dimensional moduli spaces in Ref.~\cite{Yamaguchi:2004bt} and generalized to arbitrary Calabi-Yau manifolds in Ref.~\cite{Alim:2007qj} and further studied in Refs.~\cite{Alim:2008kp,Hosono:2008ve}. This note gives an overview as well as further examples applying the construction of Ref.~\cite{Alim:2013eja}, see also \cite{Zhou:2013hpa}. Special choices of the generators of these polynomial rings are considered as well as special coordinates. The resulting special polynomial rings coincide with the known rings of quasi modular forms for cases where the duality group of the Calabi-Yau is given by a subgroup of $SL(2,\mathbbm{Z})$ and provides a generalization thereof in other cases.

%%%%%%%%%%%%%%%%%%%%%%%%%%%%%%%%%

\section{Introduction}
We start by briefly reviewing some notions of quasi modular forms.
\subsection{Modular Forms}
Let $\mathbbm{H}$ be the upper half plane
\begin{equation}
\mathbbm{H}=\{\tau \in \mathbbm{C}\, | \, \textrm{Im} \tau >0 \}\,.
\end{equation}
There is an action of $SL(2,\mathbbm{R})$ on the upper half plane given by:
\begin{equation}
\tau \mapsto \frac{a\tau + b}{c \tau +d} \,, \quad \left( \begin{array}{cc} a&b\\c&d  \end{array}\right) \in SL(2,\mathbbm{R})\,.
\end{equation}
A modular form is a meromorphic function on $\mathbbm{H}$ which has the following transformation property
\begin{equation}
f\left( \frac{a\tau +b}{c\tau+d}\right) = \left( c\tau+d\right)^k \,f(\tau)\,,\quad \left( \begin{array}{cc} a&b\\c&d  \end{array}\right) \in \Gamma \subset SL(2,\mathbbm{Z})\,,
\end{equation}
where $\Gamma$ is a discrete subgroup of $SL(2,\mathbbm{Z})$.

The group $PSL(2,\mathbbm{Z})=SL(2,\mathbbm{Z})/\mathbbm{Z}_2$ is generated by the two generators:
\begin{equation}
T=\left( \begin{array}{cc} 1&1\\0&1  \end{array}\right)\,,\quad S=\left( \begin{array}{cc} 0&1\\-1&0  \end{array}\right).
\end{equation}

Modular forms of $PSL(2,\mathbbm{Z})$ are invariant under $T$ which translates to 
\begin{equation}
f(\tau+1)=f(\tau) \,\Rightarrow f(\tau) = \sum_{i=0}^\infty \,a_i q^i\,,\quad q=\exp(2\pi i \tau)\,.
\end{equation}

The $S$ transformation 
$ \tau \mapsto -\frac{1}{\tau}$ sometimes has an interpretation of an S-duality in physical contexts, when $\tau$ is identified with a complexified coupling. Theories invariant under S-duality are exceptional and often related to $\mathcal{N}=4$ theories as in Ref.~\cite{Vafa:1994tf}. 

\begin{figure}[h!]
  \centering
  \subfloat[$PSL(2,\mathbbm{Z})$]{\includegraphics[width=0.4\textwidth]{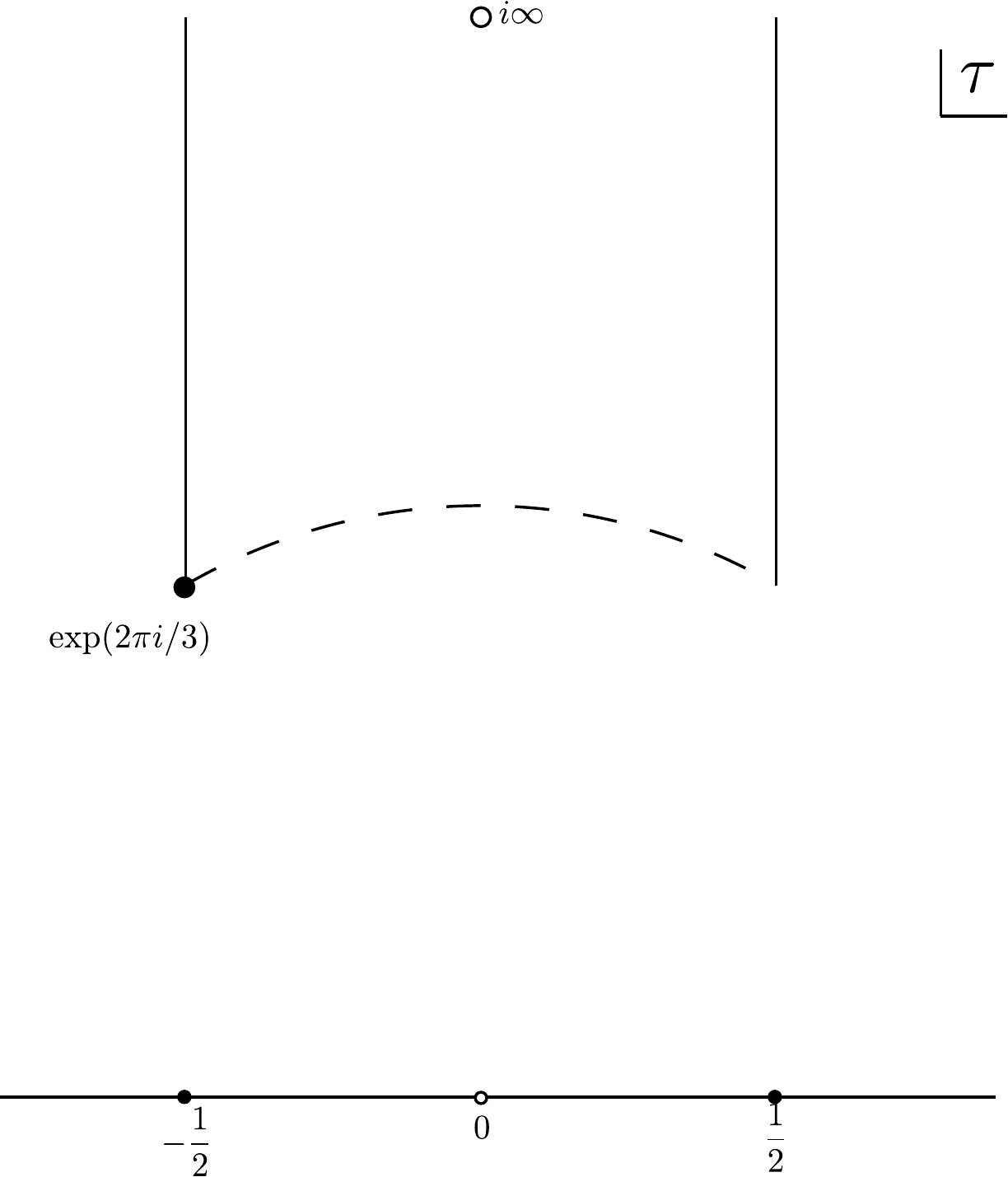}}
  \qquad
  \subfloat[$\Gamma_0(3)$]{\includegraphics[width=0.4\textwidth]{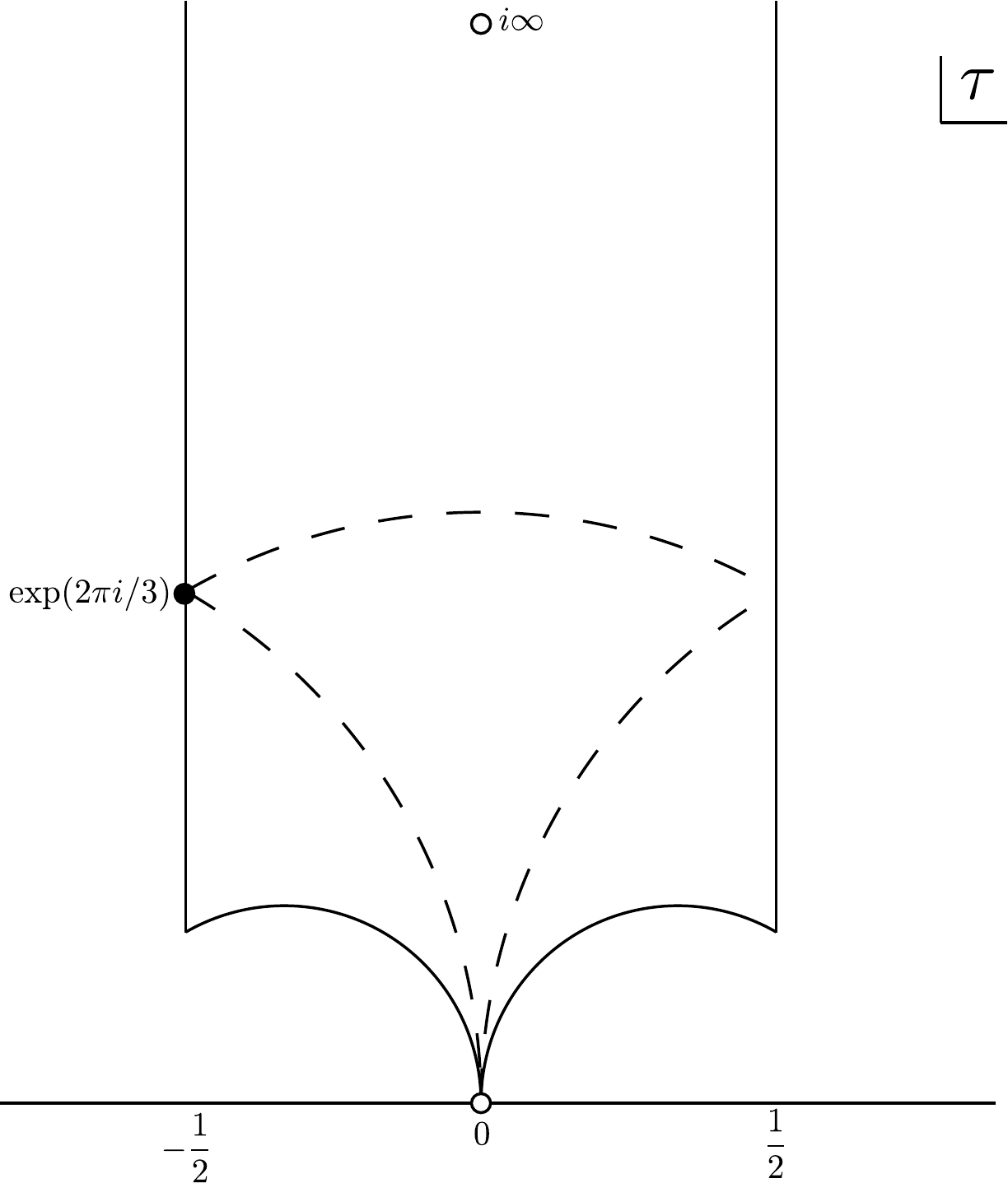}}
\caption{Fundamental domains for $PSL(2,\mathbbm{Z})$ and
$\Gamma_0(3)$. \\The empty and full circles stand for cusps and elliptic points respectively.} \label{funddomain}
\end{figure}

The quotient space $PSL(2,\mathbbm{Z})\backslash \mathbbm{H}$ is an example of a moduli space, that of inequivalent elliptic curves obtained from the quotienting $\mathbbm{C}$ by a lattice spanned by $1,\tau$. This moduli space is illustrated in Fig.~\ref{funddomain} and has one cusp at $i\infty$ which is identified with $0$ by the S-transformation. Dualities of $\mathcal{N}=2$ theories are more interesting in the sense that they correspond to the exchange of distinguished members of deformation families which do not correspond to the same point in the moduli space. An example of $\mathcal{N}=2$ duality of this kind is given in Ref.~\cite{Seiberg:1994rs}. The analog of this duality will be discussed in the following. It corresponds to the exchange of different cusps in larger moduli spaces which are obtained from subgroups of $SL(2,\mathbbm{Z})$.

%%%%%%%%%%%%%%%%%%%%%%%%%%%%%%%%%

\subsection{Congruence subgroups}

Consider the following subgroups of $SL(2,\mathbbm{Z})$
\begin{equation}
\Gamma_0(N)= \left\{ \left( \begin{array}{cc} a&b\\c&d  \end{array}\right) \in SL(2,\mathbbm{Z}) \,  | \,c \equiv 0\, \textrm{mod}\, N  \right\}\,.
\end{equation}

The Fricke involution exchanges two different cusps in the fundamental domains of these subgroups:
\begin{equation}\label{fricke}
W_N: \tau \mapsto -\frac{1}{N\tau}\,,
\end{equation}
the corresponding action on the quasi modular forms significantly facilitates obtaining the expansions of topological string amplitudes in different distinguished patches of moduli space. The fundamental domain of $\Gamma_0(3)$ is shown in Fig.~\ref{funddomain}. This group will appear in the study of local $\mathbbm{P}^2$.

%%%%%%%%%%%%%%%%%%%%%%%%%%%%%%%%%

\subsection{Differential ring of quasi modular forms}
One may ask what happens when we take a derivative of a modular form, suppose we are given the following modular form of weight 12:
\begin{equation}
\Delta(\tau)= \eta^{24}(\tau) = q\, \prod_{n=1}^\infty (1-q^n)^{24}\,.
\end{equation}
The following derivative defines a new function $E_2$ on $\mathbbm{H}$
\begin{equation}\label{defE2}
\frac{1}{2\pi i} \partial_{\tau} \log \Delta = E_2\,,
\end{equation}
we may proceed by taking derivatives and casting the outcome in the following form, which successively defines the functions $E_4$ and $E_6$:
\begin{eqnarray}
\frac{1}{2\pi i} \partial_\tau E_2= \frac{1}{12} (E_2^2-E_4)\,, \\
\frac{1}{2\pi i} \partial_\tau E_4= \frac{1}{3} (E_2\,E_4-E_6)\,, \\
\frac{1}{2\pi i} \partial_\tau E_6= \frac{1}{2} (E_2\,E_6-E_4^2)\,.
\end{eqnarray}
Something non-trivial happens in the third equation, where no new function is generated, but instead the differential ring of functions closes. $E_4$ and $E_6$ are the Eisenstein series and are modular forms of $SL(2,\mathbbm{Z})$ of weights $4$ and $6$, every higher weight modular form can be written as a polynomial in these two. The function $E_2$ does not quite transform in a modular way as can be easily checked from the defining equation (\ref{defE2}). It can be completed to a modular form by adding a non-holomorphic piece:
\begin{equation}
\widehat{E}_2 = E_2 -\frac{3}{\pi \textrm{Im}\tau}\,,
\end{equation}
$E_2$ is then called a quasi-modular form and $\widehat{E}_2$ an almost holomorphic modular form \cite{Kaneko:1995}. It is the analogous structure which can be defined using the special geometry of Calabi-Yau threefolds which is the subject of this note.

%%%%%%%%%%%%%%%%%%%%%%%%%%%%%%%%%
\section{Special geometry and holomorphic anomaly}

\subsection{Special geometry}
Let $\mathcal{M}$ be the moduli space of a Calabi-Yau threefold, this can be the moduli space of complexified K\"ahler forms of a CY $X$ or the moduli space of complex structures of the mirror $\check{X}$. $\mathcal{M}$ is a special K\"ahler manifold, which is a K\"ahler manifold whose curvature can be expressed in a special form described below. The special geometry of $\mathcal{M}$ can be best described using a bundle $\mathcal{H}\rightarrow \mathcal{M}$. Fixing a point on the base manifold, this bundle can be decomposed into subbundles in the following way:
\begin{equation}
\mathcal{H}= \mathcal{L} \oplus \left(\mathcal{L}\otimes \,T\mathcal{M}\right)  \oplus \overline{\left(\mathcal{L}\otimes \,T\mathcal{M}\right)}  \oplus \overline{\mathcal{L}}\,,
\end{equation}
where $\mathcal{L}$ is a line bundle on $\mathcal{M}$, $T\mathcal{M}$ denotes the holomorphic tangent bundle and the overline denotes complex conjugation. The change of base point on $\mathcal{M}$ changes the decomposition of the bundle $\mathcal{H}$. An example of the this structure is the variation of Hodge structure on the B-side of mirror symmetry where $\mathcal{H}=H^3(\check{X},\mathbbm{C})$ and the decomposition into subbundles is the Hodge decomposition. 

In the following we will use local coordinates $x^i\,, i=1,\dots,n=\textrm{dim}\,(\mathcal{M})$ and denote $\partial_i=\partial/\partial x^i\,, \partial_{\bar{\jmath}}=\partial/\partial \bar{x}^{\jmath}$.  $e^{-K}$ is a metric on $\mathcal{L}$ with connection $K_i$, it provides a K\"ahler form for a K\"ahler metric on $\mathcal{M}$, whose components and Levi-Civita connection are given by:
\begin{equation}
G_{i\bar{\jmath}} := \partial_i \partial_{\bar{\jmath}} K\,, \quad \Gamma_{ij}^k= G^{k\bar{k}} \partial_i G_{j\bar{k}}\,.
\end{equation}
The description of the change of the decomposition of $\mathcal{H}$ into subbundles, reminiscent of the chiral ring of CFT, is captured by the holomorphic Yukawa couplings or threepoint functions
\begin{equation}
C_{ijk} \in \Gamma\left( \mathcal{L}^2 \otimes \textrm{Sym}^3 T^*\mathcal{M}\right)\,,
\end{equation}
giving the following expression for the curvature:
 \begin{equation}
R_{i\bar{\imath}\phantom{l}j}^{\phantom{i\bar{\imath}}l}=[\bar{\partial}_{\bar{\imath}},D_i]^l_{\phantom{l}j}=\bar{\partial}_{\bar{\imath}} \Gamma^l_{ij}= \delta_i^l
G_{j\bar{\imath}} + \delta_j^l G_{i\bar{\imath}} - C_{ijk} \overline{C}^{kl}_{\bar{\imath}},
\label{curvature}
 \end{equation}
where $D_i$ denotes the covariant derivative and:
\begin{equation}
\overline{C}_{\bar{\imath}}^{jk}:= e^{2K} G^{k\bar{k}} G^{j\bar{\jmath}}\overline{C}_{\bar{\imath}\bar{k}\bar{\jmath}}.
\end{equation}
We further introduce the objects $S^{ij},S^i,S$, which are sections of $\mathcal{L}^{-2}\otimes \text{Sym}^m T\mathcal{M}$ with $m=2,1,0$, respectively, and give local potentials for the non-holomorphic Yukawa couplings:
\begin{equation}
\partial_{\bar{\imath}} S^{ij}= \overline{C}_{\bar{\imath}}^{ij}, \qquad
\partial_{\bar{\imath}} S^j = G_{i\bar{\imath}} S^{ij}, \qquad
\partial_{\bar{\imath}} S = G_{i \bar{\imath}} S^i.
\label{prop}
\end{equation}

%%%%%%%%%%%%%%%%%%%%%%%%%%%%%%%%%

\subsection{Holomorphic anomaly equations}
The topological string amplitudes at genus $g$ with $n$ insertions $\mathcal{F}^{g}_{i_1\dots i_n}$  are defined in Ref.~\cite{Bershadsky:1993cx} are sections of the line bundles
$\mathcal{L}^{2-2g}$ over $\mathcal M$. These are only non-vanishing for
$(2g-2+n)>0$. They are related recursively in $n$ by
\begin{equation}
D_i \mathcal{F}^{(g)}_{i_1\cdots i_n}=\mathcal{F}^{(g)}_{ii_1\cdots i_n},
\end{equation}
as well as in $g$ by the holomorphic anomaly equation for $g=1$ \cite{Bershadsky:1993ta}
\begin{equation}
\bar{\partial}_{\bar{\imath}} \mathcal{F}^{(1)}_j = \frac{1}{2} C_{jkl}
\overline{C}^{kl}_{\bar{\imath}}+ (1-\frac{\chi}{24})
G_{j \bar{\imath}}\,, \label{anom2}
\end{equation}
where $\chi$ is the Euler character of the CY threefold. As well as for $g>2$ \cite{Bershadsky:1993cx}:

\begin{equation}
\bar{\partial}_{\bar{\imath}} \mathcal{F}^{g} = \frac{1}{2} \overline{C}_{\bar{\imath}}^{jk} \left(
\sum_{r=1}^{g-1}
D_j\mathcal{F}^{(r)} D_k\mathcal{F}^{(g-r)} +
D_jD_k\mathcal{F}^{(g-1)} \right) \label{anom1}.
\end{equation}

%%%%%%%%%%%%%%%%%%%%%%%%%%%%%%%%%

\subsection{Feynman diagram solution}
BCOV showed that the higher genus amplitudes can be cast in terms of Feynman diagrams with propagators $S^{ij},S^i,S$ and vertices $\mathcal{F}^{g}_{i_1\dots i_n}$. To obtain that solution, the $\partial_{\bar{i}}$ derivatives are integrated on both sides of Eq.~(\ref{anom1}) using 
$$\partial_{\bar{\imath}} S^{ij}= \overline{C}_{\bar{\imath}}^{ij}, \qquad$$
and the integrated special geometry relation which can be obtained from Eq.~\ref{curvature}.
\begin{equation}
\Gamma^l_{ij} = \delta_i^l K_j + \delta^l_j K_i - C_{ijk} S^{kl} + s^l_{ij}\,,
\label{specgeom}
\end{equation}
where $s_{ij}^l$ are holomorphic functions.

%%%%%%%%%%%%%%%%%%%%%%%%%%%%%%%%%

\section{Special polynomial rings}
In the following it will be shown that using special geometry a differential ring of functions on the moduli space can be defined.
\subsection{Polynomial structure}
In Ref.~\cite{Yamaguchi:2004bt} it was shown that the topological string amplitudes can be expressed as polynomials in finitely many generators of differential ring of multi-derivatives of the connections of special geometry, it was shown that there are relations among the infinitely many generators:
\begin{equation}
(\partial_z)^p K_z\,,\quad (\partial_z)^p \Gamma_{zz}^z\,,\quad p=0,\dots,\infty,
\end{equation}
where $z$ is a local coordinate on $\mathcal{M}$. The purely holomorphic part of the amplitudes as well as the holomorphic ambiguities were expressed as rational functions in $z$.

This construction was generalized in Ref.~\cite{Alim:2007qj} for any CY manifold. It was shown there that $\mathcal{F}^g_{i_1,\dots,i_n}$ is a polynomial of degree $3g-3+n$ in the generators $S^{ij},S^i,S,K_i$ where degrees $1,2,3,1$ were assigned to these generators respectively. The purely holomorphic part of the construction as well as the coefficients of the monomials would be rational functions in the algebraic moduli. 

%%%%%%%%%%%%%%%%%%%%%%%%%%%%%%%%%

\subsection{Special coordinates}
In the following a special set of coordinates $t^a$ on $\mathcal{M}_{cplx}$, the complex structure moduli space of $\check{X}$ will be discussed. The holomorphic $(3,0)$ form $\Omega(x)$  can be expanded in terms of a symplectic basis $\alpha_I,\beta^J \in H^3(\check{X},\mathbbm{Z})\,, I,J=0,\dots n$:
\begin{equation}
\Omega(x)= X^I(x) \alpha_I + \mathcal{F}_J(x) \beta^J\, .
\end{equation}
The periods $X^I(x),\mathcal{F}_J(x)$ satisfy the Picard--Fuchs equation of the  B-model CY family. $X^I$ can be identified with projective coordinates on $\mathcal{M}$ and $\mathcal{F}_J$ with derivatives of a homogeneous function $\mathcal{F}(X^I)$ of weight 2 such that $\mathcal{F}_J=\frac{\partial \mathcal{F}(X^I)}{\partial X^J}$. In a patch where $X^0(x)\ne 0$ a set of special coordinates can be defined
\begin{equation} \label{special}
t^a=\frac{X^a}{X^0}\, ,\quad a=1,\dots ,h^{2,1}(\check{X}).
\end{equation}
The normalized holomorphic $(3,0)$ form $ (X^0)^{-1} \Omega(t)$ has the expansion:
\begin{equation}
 (X^0)^{-1} \Omega(t)= \alpha_0 + t^a \alpha_a +\beta^b F_b(t) + (2F_0(t)-t^c F_c(t)) \beta^0\,,
\end{equation}
where $$F_0(t)= (X^0)^{-2} \mathcal{F} \quad \textrm{and} \quad F_a(t):=\partial_a F_0(t)=\frac{\partial F_0(t)}{\partial t^a}.$$
$F_0(t)$ is the prepotential. We furthermore have:
\begin{equation}
C_{abc}=\partial_a \partial_b \partial_c F_0(t)\,.
\end{equation}

%%%%%%%%%%%%%%%%%%%%%%%%%%%%%%%%%

\subsection{Special rings}
We now restrict to one dimensional moduli spaces and introduce the differential rings of Ref.~\cite{Alim:2013eja}.  We consider the following polynomial generators obtained from $S^{zz},S^z,S$
\begin{eqnarray}
S^{tt}=(\partial_z t)^2 (X^0)^2\, S^{zz} \,,\quad \tilde{S}^t=(\partial_z t) (X^0)^2\, (S^z-S^{zz}K_z)\,,\nonumber\\\tilde{S}_0=(X^0)^2\, (S-S^z K_z+\frac{1}{2}S^{zz}K_z)\,, \quad K_t= (\partial_z t)^{-1} \, K_z\,,
\end{eqnarray}

Furthermore, a new coordinate is introduced on the moduli space:
\begin{equation}
\tau=\frac{1}{\kappa} \partial_t F_t \,,
\end{equation}
where $\kappa$ is the classical triple intersection of $X$.
We define the following functions on the moduli space:
  \begin{equation}
    \begin{aligned}
      K_0 &= \kappa\, C_{ttt}^{-1} \, (\theta t)^{-3} \,,&  G_1&=\theta t\,,& K_2&=\kappa\,C_{ttt}^{-1}K_t\,,\nonumber\\
      T_2&=S^{tt}\,,& T_4&= C_{ttt}^{-1} \tilde{S}^t\,,&
      T_6&=C_{ttt}^{-2} \tilde{S}_0\,,
    \end{aligned}
\end{equation}
where $\theta:=z \partial_z$.

The differential ring found in Ref.~\cite{Alim:2007qj} can be translated to this choice of generators and coordinates giving the following ring \cite{Alim:2013eja}:
\begin{equation}
\label{relspecial}
\begin{aligned}
\partial_{\tau} K_0&=-2K_0\,K_2- K_0^2\, G_1^2\,(\tilde{h}^z_{zzz}+3(s_{zz}^z+1))\,,\\
\partial_{\tau} G_1&= 2G_1\,K_2-\kappa  G_1\,T_2\,+K_0 G_1^3(s_{zz}^z+1)\,,\\
\partial_{\tau} K_2&=3K_2^2-3\kappa K_2\,T_2-\kappa^2 T_4+K_0^2\,G_1^4 k_{zz}-K_0\,G_1^2\,K_2\,\tilde{h}^z_{zzz}\,,\\
\partial_{\tau} T_2&=2K_2\,T_2-\kappa T_2^2+2\kappa T_4+\frac{1}{\kappa}\,K_0^2 G_1^4 \tilde{h}^z_{zz}\,,\\
\partial_{\tau} T_4&=4 K_2 T_4-3\kappa T_2\,T_4+ 2\kappa T_6-K_0\, G_1^2 \, T_4 \tilde{h}^z_{zzz}-\frac{1}{\kappa} K_0^2\, G_1^4 \,T_2 k_{zz}+\frac{1}{\kappa^2} K_0^3\, G_1^6 \tilde{h}_{zz}\,,\\
\partial_{\tau} T_6&= 6 K_2\, T_6-6\kappa T_2 \,T_6+\frac{\kappa}{2} T_4^2-\frac{1}{\kappa} K_0^2\, G_1^4 \,T_4\,k_{zz}+\frac{1}{\kappa^3} K_0^4\, G_1^8 \tilde{h}_z-2 \, K_0\,G_1^2\,T_6  \tilde{h}^z_{zzz}\,.
\end{aligned}
\end{equation}
An assumption which is confirmed in all studied examples is that there are choices of the generators such that the functions $\tilde{h}^{z}_{zz},\tilde{h}_{zz},\tilde{h}_z$ and $\tilde{h}^z_{zzz}$ and $k_{zz}$ can be expressed as rational functions in the algebraic modulus $z$. As a generator of the rational functions we consider:
\begin{equation}\label{ratgen}
C_0=\theta \log z^3 C_{zzz}\,,
\end{equation}
the derivative of this generator is computed to be:
\begin{equation}
\partial_{\tau} C_0= K_0\, G_1^2\, C_0\,(C_0+1)\,.
\end{equation}
The holomorphic anomaly equations in the polynomial formulation for  $F^{(g)}= (X^0)^{2g-2}\,\mathcal{F}^{g}\,$ now become \cite{Alim:2007qj,Alim:2013eja}:
\begin{equation}\label{specialrec1}
\frac{\partial F^{(g)}}{\partial T_2}-\frac{1}{\kappa}\frac{\partial F^{(g)}}{\partial T_4} \,K_2 +\frac{1}{2\kappa^2} \frac{\partial F^{(g)}}{\partial T_6} K_2^2 =\frac{1}{2} \sum_{r=1}^{g-1} \partial_t F^{(g-r)}\,\partial_t\, F^{(r)} +\frac{1}{2} \partial_t^2 F^{(g-1)}\,,
\end{equation}
and
\begin{equation}\label{specialrec2}
\frac{\partial F^{(g)}}{\partial K_2}=0\,.
\end{equation}
The $t$ derivative in Eq.~(\ref{specialrec1}) can be replaced by:
\begin{equation}
\partial_t= K_0^{-1} G_1^{-3} \, \partial_{\tau}\,.
\end{equation}

 $F^{(g)}$ is a polynomial of degree zero in the generators, obtained recursively from Eqs.~(\ref{specialrec1},\,\ref{specialrec2}) up to the addition of a rational function of the form
$$ A^{(g)}= K_0^{g-1}\, P^{(g)}(C_0),$$
where $P^{(g)}$ is a rational function in $C_0$ chosen such that $F^{(g)}$ respects the boundary conditions.

%%%%%%%%%%%%%%%%%%%%%%%%%%%%%%%%%

\section{Examples}
\subsection{Local $\mathbbm{P}^2$}

To fix the special polynomial ring we choose the following rational functions in $z$ in the construction of the ring (\ref{relspecial})\footnote{Multiplying (dividing) lower(upper) indices by $z$.}
\begin{equation}
s_{zz}^z=-\frac{4}{3}+\frac{1}{6\Delta}\,,\quad \tilde{h}^z_{zz}=\frac{1}{36\Delta^2}\,,\quad \tilde{h}^z_{zzz}=\frac{1}{2\Delta}\,,
\end{equation}
with $\Delta=1+27z$. The generators are $T_2,G_{1}$ and $C_0$ these can be expressed in terms of the generators of the ring of quasi modular forms of $\Gamma_0(3)$:
\begin{equation}
\begin{aligned}
&T_2=\frac{1}{8}(E_2(\tau)+3 E_2(3\tau)),  \\
&G_1=\theta_2(2\tau) \theta_3(6\tau)+\theta_2(6\tau)\theta_3(2\tau) \\ 
&C_0=27 \left(\frac{\eta(3\tau)}{\eta(\tau)}\right)^{12}.
\end{aligned}
\end{equation}
We obtain the following ring \cite{Alim:2013eja}
\begin{align}
&  \partial_{\tau} C_0= G_1^2\,C_0\,,\\
&  \partial_{\tau} G_1=\frac{1}{6}\left(2G_1\, T_2+G_1^3\left(\frac{C_0-1}{C_0+1}\right)\right)\,,\\
& \partial_{\tau} T_2=\frac{T_2^2}{3}-\frac{G_1^4}{12}\,.
\end{align}

\subsubsection*{\it Genus 1}
The genus 1 amplitude is found to be
\begin{equation}
\begin{aligned}
&F^{(1)}=-\frac{1}{12} \log\left((\theta t)^6 z(1+27z) \right)= -\frac{1}{12} \log\left(\frac{C_0\, G_1^6}{27(1+C_0)^2} \right)\,\\
&\partial_t \,F^{(1)}=-\frac{1}{6} (1+C_0) G_1^{-3} \, T_2\,.
\end{aligned}
\end{equation}

\subsubsection*{Duality action and conifold expansion}
The Fricke involution (\ref{fricke}) on this choice of generators becomes:
\begin{equation}
T_2\rightarrow 3 \tau^2 \,T_2\,, \quad G_1 \rightarrow -i \sqrt{3}\, \tau\,G_1\,, \quad C_0 \rightarrow \frac{1}{C_0}\,.
\end{equation}
Using this duality transformation allows us to obtain the expansion of $F^{(g)}$ in terms of the flat coordinate around the conifold singularity \cite{Alim:2013eja}. In terms of the latter the gap condition can be imposed to solve for the holomorphic ambiguity. This was first used in physically motivated example in Ref.~\cite{Huang:2006si}. For example at genus 2 we find \cite{Alim:2013eja}:
\begin{equation}
F^{(2)}=\frac{ (1+C_0)^2 \, T_2 \left( 3\,G_1^4-9G_1^2\, T_2 + 10 T_2^2\right)}{432 \,G_1^6} -\frac{11}{17280}-\frac{7}{4320}C_0-\frac{1}{1080} C_0^2\,.
\end{equation}

%%%%%%%%%%%%%%%%%%%%%%%%%%%%%%%%%

\subsection{Compact geometries}
In the following we give examples of the special polynomial rings for compact geometries. We consider the geometries studied in Ref.~\cite{Klemm:1992tx}. These are the quintic in $\mathbbm{P}^4$, which is the classical example of mirror symmetry of a compact geometry \cite{Candelas:1990rm}, as well as the sextic, octic and dectic in the weighted projective spaces $\mathbbm{WP}_{2,1,1,1,1}, \mathbbm{WP}_{4,1,1,1,1}$ and $\mathbbm{WP}_{5,2,1,1,1}$ respectively and their mirror geometries. The polynomial rings for these geometries were considered in Ref.~\cite{Yamaguchi:2004bt} and were used in Ref.~\cite{Huang:2006hq} for higher genus computations. The genus 0 data is given in Ref.~\cite{Klemm:1992tx}. The Yukawa coupling is given by \footnote{For the mirror manifolds in terms of an algebraic coordinate on the moduli space, and adopting the convention of multiplying lower tensorial indices by $z$.}
\begin{equation}
C_{zzz}=\frac{\kappa}{\Delta}\,,\quad \Delta=(1-\alpha z)\,.
\end{equation}
We choose as generators of the rational holomorphic functions in the algebraic modulus:
\begin{equation}
C_0=\frac{\alpha z}{1-\alpha z}\,,
\end{equation}

We fix the holomorphic functions appearing in (\ref{relspecial}) which completely fixes the choice of the generators of the differential ring. We summarize the data for all four geometries in the following, for the quintic these choices were considered in Refs.~\cite{Hosono:2008ve,Alim:2012gq,Alim:2013eja}, the choices made for the other geometries is similar:

\begin{equation}
\begin{array}{|c|c|c|c|c|}
\hline
&\textrm{Quintic}&\textrm{Sextic}&\textrm{Octic}&\textrm{Dectic}\\
\hline
\kappa&5&3&2&1\\
\alpha& 3125&11664&65536&800000\\
s_{zz}^z&-\frac{8}{5}&-\frac{3}{2}&-\frac{3}{2}&-\frac{7}{5}\\
\tilde{h}_{zz}^{z}&\frac{1}{5\Delta}&\frac{5}{36\Delta} &\frac{5}{32\Delta}&\frac{1}{10\Delta}\\
\tilde{h}_{zz}&-\frac{1}{25\Delta}&0&0&0\\
\tilde{h}_z&\frac{2}{625 \Delta}& \frac{1}{648\Delta}&\frac{9}{8192\Delta}&\frac{9}{20000\Delta}\\
k_{zz}&\frac{2}{25}&\frac{1}{18}&\frac{3}{64}&\frac{3}{100}\\
\hline
\end{array}
\end{equation}

%%%%%%%%%%%%%%%%%%%%%%%%%%%%%%%%%

\subsubsection{Higher genus amplitudes}
The initial correlation functions at genus $1$ are found.\footnote{Subscripts  Q,S,O,D are used to denote the quintic, sextic, dectic and octic}. Starting from these and using the boundary conditions as in Ref.~\cite{Huang:2006hq} higher genus expressions are obtained using the boundary conditions. These lengthy expressions will be omitted.
\begin{eqnarray}
(F^{(1)}_t)_Q&=& \frac{(5 C_0-107) G_1^2 K_0+560 K_2+150
   T_2}{60 G_1^3 K_0}\,, \\
   (F^{(1)}_t)_S&=&   \frac{(C_0-18) G_1^2 K_0+6 (19 K_2+3
   T_2)}{12 G_1^3 K_0} \,,   \\
   (F^{(1)}_t)_O&=& \frac{(C_0-19) G_1^2 K_0+4 (40 K_2+3
   T_2)}{12 G_1^3 K_0}  \,,  \\
(F^{(1)}_t)_D   &=&\frac{(5 C_0-73) G_1^2 K_0+30 (26
   K_2+T_2)}{60 G_1^3 K_0}\,.
\end{eqnarray}

%%%%%%%%%%%%%%%%%%%%%%%%%%%%%%%%%

\section{Conclusions}
In this note, an overview was given of the construction of special differential polynomial rings on the moduli spaces of CY threefolds. In special cases, these coincide with known rings of quasi modular forms, this was illustrated in the example of local $\mathbbm{P}^2$, further examples of local del Pezzo geometries can be found in Ref.~\cite{Alim:2013eja} based on Ref.~\cite{Lerche:1996ni}. The appearance of quasi modular forms in higher genus topological string theory has been a recurrent theme of many works including Refs.~\cite{Bershadsky:1993cx,Hosono:1999qc,Hosono:2002xj,Klemm:2004km,Aganagic:2006wq,Alim:2012ss}. The construction of the polynomial rings from special geometry provides a general setting which includes the known appearances of quasi modular forms and provides a generalization thereof otherwise. For compact geometries, including the quintic, these differential rings should provide strong hints of a more general theory generalizing the classical quasi modular forms, this is also subject of Ref.~\cite{Movasati:2011zz} and subsequent works. Another question that begs for an explanation is the enumerative content of the Fourier expansion coefficients of the modular forms, which is different from the expansions giving GW invariants, for some local geometries this has been addressed for example in Ref.~\cite{Stienstra}.

%    Bibliographies can be prepared with BibTeX using amsplain,
%    amsalpha, or (for "historical" overviews) natbib style.
%\bibliography{duality.bib}

\begin{thebibliography}{10}

\bibitem{Aganagic:2006wq}
Mina Aganagic, Vincent Bouchard, and Albrecht Klemm, \emph{{Topological Strings
  and (Almost) Modular Forms}}, Commun.Math.Phys. \textbf{277} (2008),
  771--819.

\bibitem{Alim:2012gq}
Murad Alim, \emph{{Lectures on Mirror Symmetry and Topological String Theory}},
   (2012).

\bibitem{Alim:2007qj}
Murad Alim and Jean~Dominique L{\"a}nge, \emph{{Polynomial Structure of the
  (Open) Topological String Partition Function}}, JHEP \textbf{0710} (2007),
  045.

\bibitem{Alim:2008kp}
Murad Alim, Jean~Dominique L{\"a}nge, and Peter Mayr, \emph{{Global Properties
  of Topological String Amplitudes and Orbifold Invariants}}, JHEP
  \textbf{1003} (2010), 113.

\bibitem{Alim:2012ss}
Murad Alim and Emanuel Scheidegger, \emph{{Topological Strings on Elliptic
  Fibrations}},  (2012).

\bibitem{Alim:2013eja}
Murad Alim, Emanuel Scheidegger, Shing-Tung Yau, and Jie Zhou, \emph{{Special
  Polynomial Rings, Quasi Modular Forms and Duality of Topological Strings}},
  (2013).

\bibitem{Bershadsky:1993ta}
M.~Bershadsky, S.~Cecotti, H.~Ooguri, and C.~Vafa, \emph{{Holomorphic anomalies
  in topological field theories}}, Nucl.Phys. \textbf{B405} (1993), 279--304.

\bibitem{Bershadsky:1993cx}
\bysame, \emph{{Kodaira-Spencer theory of gravity and exact results for quantum
  string amplitudes}}, Commun.Math.Phys. \textbf{165} (1994), 311--428.

\bibitem{Candelas:1990rm}
Philip Candelas, Xenia~C. De~La~Ossa, Paul~S. Green, and Linda Parkes, \emph{{A
  Pair of Calabi-Yau manifolds as an exactly soluble superconformal theory}},
  Nucl.Phys. \textbf{B359} (1991), 21--74.

\bibitem{Hosono:1999qc}
S.~Hosono, M.H. Saito, and A.~Takahashi, \emph{{Holomorphic anomaly equation
  and BPS state counting of rational elliptic surface}}, Adv.Theor.Math.Phys.
  \textbf{3} (1999), 177--208.

\bibitem{Hosono:2002xj}
Shinobu Hosono, \emph{{Counting BPS states via holomorphic anomaly equations}},
  Fields Inst.Commun. (2002), 57--86.

\bibitem{Hosono:2008ve}
\bysame, \emph{{BCOV ring and holomorphic anomaly equation}},  (2008).

\bibitem{Huang:2006si}
Min-xin Huang and Albrecht Klemm, \emph{{Holomorphic Anomaly in Gauge Theories
  and Matrix Models}}, JHEP \textbf{0709} (2007), 054.

\bibitem{Huang:2006hq}
Min-xin Huang, Albrecht Klemm, and Seth Quackenbush, \emph{{Topological string
  theory on compact Calabi-Yau: Modularity and boundary conditions}},
  Lect.Notes Phys. \textbf{757} (2009), 45--102.

\bibitem{Kaneko:1995}
Masanobu Kaneko and Don Zagier, \emph{A generalized {J}acobi theta function and
  quasimodular forms}, The moduli space of curves ({T}exel {I}sland, 1994),
  Progr. Math., vol. 129, Birkh\"auser Boston, Boston, MA, 1995, pp.~165--172.
  \MR{1363056 (96m:11030)}

\bibitem{Klemm:2004km}
A.~Klemm, M.~Kreuzer, E.~Riegler, and E.~Scheidegger, \emph{{Topological string
  amplitudes, complete intersection Calabi-Yau spaces and threshold
  corrections}}, JHEP \textbf{0505} (2005), 023.

\bibitem{Klemm:1992tx}
Albrecht Klemm and Stefan Theisen, \emph{{Considerations of one modulus
  Calabi-Yau compactifications: Picard-Fuchs equations, Kahler potentials and
  mirror maps}}, Nucl.Phys. \textbf{B389} (1993), 153--180.

\bibitem{Lerche:1996ni}
W.~Lerche, P.~Mayr, and N.P. Warner, \emph{{Noncritical strings, Del Pezzo
  singularities and Seiberg-Witten curves}}, Nucl.Phys. \textbf{B499} (1997),
  125--148.

\bibitem{Movasati:2011zz}
Hossein Movasati, \emph{{Eisenstein type series for Calabi-Yau varieties}},
  Nucl.Phys. \textbf{B847} (2011), 460--484.

\bibitem{Seiberg:1994rs}
N.~Seiberg and Edward Witten, \emph{{Electric - magnetic duality, monopole
  condensation, and confinement in N=2 supersymmetric Yang-Mills theory}},
  Nucl.Phys. \textbf{B426} (1994), 19--52.

\bibitem{Stienstra}
Jan Stienstra, \emph{Mahler measure, {E}isenstein series and dimers}, Mirror
  symmetry. {V}, AMS/IP Stud. Adv. Math., vol.~38, Amer. Math. Soc.,
  Providence, RI, 2006, pp.~151--158. \MR{2282959 (2008c:11120)}

\bibitem{Vafa:1994tf}
Cumrun Vafa and Edward Witten, \emph{{A Strong coupling test of S duality}},
  Nucl.Phys. \textbf{B431} (1994), 3--77.

\bibitem{Yamaguchi:2004bt}
Satoshi Yamaguchi and Shing-Tung Yau, \emph{{Topological string partition
  functions as polynomials}}, JHEP \textbf{0407} (2004), 047.

\bibitem{Zhou:2013hpa}
Jie Zhou, \emph{{Differential Rings from Special K\"ahler Geometry}},  (2013).

\end{thebibliography}
\bibliographystyle{amsplain}
%    Insert the bibliography data here.

\providecommand{\bysame}{\leavevmode\hbox to3em{\hrulefill}\thinspace}
\providecommand{\MR}{\relax\ifhmode\unskip\space\fi MR }
% \MRhref is called by the amsart/book/proc definition of \MR.
\providecommand{\MRhref}[2]{%
  \href{http://www.ams.org/mathscinet-getitem?mr=#1}{#2}
}
\providecommand{\href}[2]{#2}

\end{document}